\documentclass[iop,apj]{emulateapj}
\pdfoutput=1
\usepackage{apjfonts}
\usepackage{amsmath,amssymb}
\usepackage{xspace}
\usepackage{url}
\usepackage{datetime}
\usepackage[backref, breaklinks,colorlinks, urlcolor=magenta, citecolor=blue, linkcolor=red]{hyperref}

\usepackage[all]{hypcap} %Links go to figures;breaks on deluxetables
\slugcomment{Accepted for publication in The Astrophysical Journal}

\shorttitle{Stellar populations in massive quenched galaxies at $z\simeq1.6$}
\shortauthors{Onodera et al.}

\newcommand{\kms}{\ensuremath{\,\text{km}~\text{s}^{-1}}}
\newcommand{\hda}{H$\delta_\mathrm{A}$}
\newcommand{\hdf}{H$\delta_\mathrm{F}$}
\newcommand{\hga}{H$\gamma_\mathrm{A}$}
\newcommand{\hgf}{H$\gamma_\mathrm{F}$}
\newcommand{\hbeta}{H$\beta$}
\newcommand{\cnone}{CN$_1$}
\newcommand{\cntwo}{CN$_2$}
\newcommand{\gband}{G4300}
\newcommand{\mgone}{Mg$_1$}
\newcommand{\mgtwo}{Mg$_2$}
\newcommand{\mgb}{Mg\textit{b}}
\newcommand{\ctwo}{C$_2$4668}
\newcommand{\mz}{\ensuremath{\langle z\rangle}}
\newcommand{\mm}{\ensuremath{\langle M\rangle}}
\newcommand{\sersic}{S\'ersic\xspace}
\newcommand{\ppxf}{pPXF\xspace}
\newcommand{\zoh}{\ensuremath{\text{[Z/H]}}\xspace}
\newcommand{\feh}{\ensuremath{\text{[Fe/H]}}\xspace}
\newcommand{\afe}{\ensuremath{\text{[$\alpha$/Fe]}}\xspace}
\newcommand{\logage}{\ensuremath{\log(\text{age})}\xspace}

\begin{document}

\title{The ages, metallicities and element abundance ratios of massive quenched galaxies at $z\simeq1.6^{1}$}

\author{
  M. Onodera\altaffilmark{2},
  C.~M. Carollo\altaffilmark{2},
  A. Renzini\altaffilmark{3},
  M. Cappellari\altaffilmark{4},
  C. Mancini\altaffilmark{3,5},
  N. Arimoto\altaffilmark{6,7,8},
  E. Daddi\altaffilmark{9},
  R. Gobat\altaffilmark{9,10},
  V. Strazzullo\altaffilmark{9,11},
  S. Tacchella\altaffilmark{2},
  Y. Yamada\altaffilmark{8}
}

\altaffiltext{1}{
  Based on data collected at the Subaru telescope,
  which is operated by the National Astronomical Observatory of Japan.
  (Proposal IDs: S09A-043, S10A-058, and S11A-075.)
}

\affil{$^{2}$Institute for Astronomy, ETH Z\"urich, Wolfgang-Pauli-strasse 27, 8093 Z\"urich, Switzerland}
\affil{$^{3}$INAF-Osservatorio Astronomico di Padova, Vicolo dell'Osservatorio 5, I-35122, Padova, Italy}
\affil{$^{4}$Sub-department of Astrophysics, Department of Physics, University of Oxford, Denys Wilkinson Building, Keble Road, Oxford OX1 3RH, UK}
\affil{$^{5}$Dipartimento di Fisica e Astronomia di Padova, Universit\'a di Padova, Vicolo dell'Osservatorio, 3, I-35122, Padova, Italy}
\affil{$^{6}$Graduate University for Advanced Studies, 2-21-1 Osawa, Mitaka, Tokyo, Japan}
\affil{$^{7}$Subaru Telescope, 650 North A'ohoku Place, Hilo, Hawaii 96720, USA}
\affil{$^{8}$National Astronomical Observatory of Japan, 2-21-1 Osawa, Mitaka, Tokyo, Japan}
\affil{$^{9}$CEA, Laboratoire AIM-CNRS-Universit\'e Paris Diderot, Irfu/SAp, Orme des Merisiers, F-91191 Gif-sur-Yvette, France}
\affil{$^{10}$School of Physics, Korea Institute for Advanced Study, Heogiro 85, Seoul 130-722, Republic of Korea}
\affil{$^{11}$Department of Physics, Ludwig-Maximilians-Universit\"at, Scheinerstr. 1, D-81679 M\"unchen, Germany}

\email{monodera@phys.ethz.ch}

\begin{abstract}
We investigate the stellar population properties of 
a sample of 24 massive quenched galaxies at $1.25<z_\text{spec}<2.09$ 
identified in the COSMOS field with our Subaru/MOIRCS near-IR spectroscopic observations. 
Tracing the stellar population properties 
as close to their major formation epoch as possible, 
we try to put constraints on the star formation history, 
post-quenching evolution, and possible progenitor star-forming populations
for such massive quenched galaxies. 
By using a set of Lick absorption line indices on a rest-frame optical composite spectrum, 
the average age, metallicity \zoh, and $\alpha$-to-iron element abundance ratio \afe 
are derived as 
$\log(\text{age}/\text{Gyr})=0.04_{-0.08}^{+0.10}$, 
$\zoh=0.24_{-0.14}^{+0.20}$, and $\afe=0.31_{-0.12}^{+0.12}$, respectively. 
If our sample of quenched galaxies at $\mz=1.6$ is evolved passively to $z=0$,
their stellar population properties will align in excellent agreement 
with local counterparts at similar stellar velocity dispersions, 
which qualifies them as progenitors of local massive early-type galaxies. 
Redshift evolution of stellar population ages in quenched galaxies 
combined with low redshift measurements from the literature 
suggests a formation redshift of $z_\text{f} \sim 2.3$ 
around which the bulk of stars in these galaxies have been formed. 
The measured \afe value indicates a star formation timescale of $\lesssim 1$ Gyr, 
which can be translated into a specific star formation rate of $\simeq 1\,\text{Gyr}^{-1}$ 
prior to quenching. 
Based on these findings, 
we discuss identifying possible progenitor star-forming galaxies at $z \simeq 2.3$. 
We identify normal star-forming galaxies, i.e, those on the star-forming main sequence, 
followed by a rapid quenching event, as likely precursors of the quenched galaxies at $\mz=1.6$ presented here.
\end{abstract}


\section{Introduction}
Galaxies at any cosmic epoch appear to spend most of their active phases 
on the so-called ``main sequence'' of star-forming galaxies in which the star formation
rate (SFR) is tightly correlated with the galaxy stellar mass ($M$),
with a scatter of only $\sim 0.3$ dex 
\citep[e.g.,][]{brinchmann:2004, daddi:2007:sfr, elbaz:2007, noeske:2007:ms, pannella:2009, pannella:2014, magdis:2010, peng:2010, karim:2011, rodighiero:2011,rodighiero:2014, salmi:2012, whitaker:2012:ms,whitaker:2014:ms,kashino:2013, speagle:2014}.
In several of these studies the star-forming main sequence has a slope that is nearly independent
of redshift, while in others it evolves appreciably.
However, all studies agree that the normalization evolves strongly,
and compared to the present-day universe, is a factor of $\sim 20$ higher in SFRs with a given stellar mass near $z\sim2$, 
with a nearly constant scatter of $\sim 0.3$ dex. This suggests that 
star formation in galaxies on the main sequence is  
self-regulated in a quasi-steady state, being fed by the accretion
of cold gas coming in and powering galactic winds going out
\citep[][see \citealt{kelson:2014} for a different interpretation of the star-forming main sequence]{dekel:2009:nature,bouche:2010,dave:2012,lilly:2013}. 

Passively evolving galaxies that show fully quenched or barely detectable  amounts of star-forming
activity have also been identified out to $z\sim2$ and beyond
\citep[e.g.,][]{cimatti:2004,cimatti:2008, glazebrook:2004, daddi:2005:pbzk, kriek:2008:survey, onodera:2010:pbzk, onodera:2012, vandesande:2011, vandesande:2013, toft:2012, bedregal:2013, bezanson:2013, gobat:2013, whitaker:2013, belli:2014a, belli:2014b, krogager:2014},
including one at $z=3$ \citep{gobat:2012}.  
Various physical processes responsible for quenching star formation in star-forming main-sequence galaxies 
are currently considered and debated, 
such as the suppression of cold gas streams in dark matter halos above a critical mass \citep{birnboim:2003}, 
radio-mode \citep{croton:2006}, or quasar-mode \citep{hopkins:2006} feedback from active galactic nuclei (AGNs), 
and the suppression of disk instabilities against fragmentation into massive star-forming clumps 
\citep[so-called ``morphological'' or  ``gravitational'' quenching,][]{martig:2009,genzel:2014a}.

Fossil records imprinted in the stellar populations of quenched galaxies
(i.e., age, metallicity \zoh, and $\alpha$-element-to-iron abundance ratio \afe) 
give clues as to when quenching took place, 
and the timescales over which the stellar mass built up. 
Extensive studies have been carried out to measure these quantities in local quenched galaxies, especially
using a set of absorption line strength indicators known as the Lick indices 
\citep[e.g.,][]{burstein:1984,carollo:1993,worthey:1994,worthey:1997,trager:1998,trager:2000a,trager:2008,thomas:2005,thomas:2010,sanchezblazquez:2006:ii,yamada:2006,kuntschner:2010,spolaor:2010, greene:2012, greene:2013}. 
These studies indicate that in the local Universe the most massive ($M>10^{11}M_\odot$) among the
quenched galaxies are the oldest, most metal-rich, and most
$\alpha$-element enhanced, i.e., they typically host $\simeq 10$ Gyr
old stellar populations, with solar or above solar metallicities and
enhanced \afe ratios up to about 0.5 dex compared to the solar ratio.
The enhanced \afe ratio favors a dominance of Type II supernovae (SNe II) in chemical
enrichment relative to SNe Ia, which in turn implies  a short
timescale for star formation and quenching \citep[e.g.,][]{matteucci:1986,thomas:1999}. 

Stellar population ages of the quenched galaxy population 
have been used as an indicator of the formation epoch when the bulk of stars were formed \citep[e.g.,][]{kelson:2001}. 
These ages are derived via the strengths of Balmer absorption lines such as H$\beta$, H$\gamma$, and H$\delta$ 
that change rapidly in stellar populations younger than several Gyr,
while the changes are much milder for older ages.  
For instance, stellar population synthesis models by \citet{thomas:2011:tmj} predict 
that for solar metallicity and element abundance ratios,
the Lick H$\beta$ index changes by $\sim 2$ \AA{}  for ages between 1 and 5 Gyr,
while the change becomes only $\simeq 0.3$ \AA{} between 5 and 10 Gyr. 
The reduced age sensitivity of the H$\beta$ index at older ages can then make it difficult to estimate the formation redshift of local quenched galaxies, 
even if very high signal-to-noise ratios (S/N), e.g., $\gg 100$, can be achieved for such nearby objects \citep[e.g.,][]{yamada:2006,conroy:2014}. 
At higher redshifts, where the galaxies are closer to their quenching epoch, the Lick Balmer-line indices are more sensitive to age.
Therefore the formation redshift of quenched galaxies and of their precursors can be more precisely estimated. 

Stellar population analyses based on the Lick spectral indices have so far reached intermediate redshifts  ($z\sim0.9$),
indicating  high metallicities, \afe ratios for the most massive quenched systems, similar to those of local quenched galaxies of the same mass, and a formation redshift of $z_\text{f}>2$ \citep[][see also \citealt{choi:2014,gallazzi:2014} for the results based on different approaches]{kelson:2006,jorgensen:2013}.

In this paper we construct a composite spectrum by stacking  the spectra of 24 quenched galaxies at $1.25<z<2.09$ 
taken with Multi-object Infrared Camera and Spectrograph \citep[MOIRCS;][]{ichikawa:2006:moircs,suzuki:2008:moircs}
on the Subaru telescope, thus 
pushing the  measurement of  stellar population parameters via Lick indices
out to an average redshift of $\mz = 1.6$.  
Using the measured properties in galaxies still close to their quenching epoch
we derive more precise estimates for their formation timescale and time elapsed since their quenching,
and identify their possible  star-forming precursors. 

The paper is organized as follows.  
In \hyperref[sec:sample]{Section~\ref*{sec:sample}}, we present the sample galaxies and their composite spectrum.  
The measurement of Lick indices and the derivation of stellar population parameters (age, \zoh, and \afe) are described
in \hyperref[sec:lick_measurement]{Section~\ref*{sec:lick_measurement}} and 
\hyperref[sec:stellarpop]{Section~\ref*{sec:stellarpop}}, respectively.
In \hyperref[sec:discussion]{Section~\ref*{sec:discussion}}
we first discuss a possible selection bias in our spectroscopic sample, 
and then discuss the star formation history (SFH) of the quenched galaxies at $\mz=1.6$, along with their possible descendants and precursors 
at different redshifts.
Finally, we summarize our results in 
\hyperref[sec:summary]{Section~\ref*{sec:summary}}.
Throughout the analysis, we adopt a $\Lambda$-dominated cold dark matter ($\Lambda$CDM) cosmology 
with cosmological parameters of
$H_0=70\kms\,\mathrm{Mpc}^{-1}$, $\Omega_{\rm m}=0.3$, and $\Omega_\Lambda=0.7$ 
and AB magnitude system \citep{oke:1983}.

%%%%%%%%%%%%%%%%%%%%%%%%%%%%%%%%%%%%%%%%%%%%%%%%%%%%%%%%%%%

\section{Data}
\label{sec:sample}

\capstartfalse 
\begin{deluxetable*}{ccccccc}
  \tablewidth{0pt}
  \tablecolumns{11}
  \tablecaption{Properties of the sample galaxies \label{tab:objprop}}
  \tablehead{
    \colhead{ID} &
    \colhead{R.A.} &
    \colhead{Decl.} &
    \colhead{$z_\text{spec}$} &
    \colhead{$\log M/M_\odot$} &
    \colhead{$r_\text{e}$} &
    \colhead{\sersic $n$}
    \\
    \colhead{} &
    \colhead{(deg)} &
    \colhead{(deg)} &
    \colhead{} &
    \colhead{} &
    \colhead{(kpc)} &
    \colhead{}
  }
  \startdata
254025 & 150.6187115 & 2.0371363 & $1.8228\pm 0.0006$ & $11.64_{-0.03}^{+0.15}$  & $3.16 \pm 0.61$ & $3.4 \pm 0.4$ \\
217431 & 150.6646939 & 1.9497545 & $1.4277\pm 0.0015$ & $11.82_{-0.15}^{+0.03}$  & $7.19 \pm 1.95$ & $3.8 \pm 0.6$ \\
307881 & 150.6484873 & 2.1539903 & $1.4290\pm 0.0009$ & $11.75_{-0.11}^{+0.03}$  & $2.68 \pm 0.12$ & $2.3 \pm 0.1$ \\
233838 & 150.6251048 & 1.9889180 & $1.8199\pm 0.0016$ & $11.64_{-0.25}^{+0.16}$  & $2.25 \pm 0.31$ & $3.1 \pm 0.3$ \\
277491 & 150.5833512 & 2.0890266 & $1.8163\pm 0.0038$ & $11.56_{-0.12}^{+0.11}$  & $2.46 \pm 0.11$ & $1.0$\tablenotemark{a} \\
250093 & 150.6053729 & 2.0288998 & $1.8270\pm 0.0010$ & $11.25_{-0.08}^{+0.13}$  & $3.00 \pm 0.15$ & $1.0$\tablenotemark{a} \\
263508 & 150.5677283 & 2.0594318 & $1.5212\pm 0.0009$ & $11.11_{-0.21}^{+0.10}$  & $0.86 \pm 0.03$ & $3.2 \pm 0.2$ \\
269286 & 150.5718552 & 2.0712204 & $1.6593\pm 0.0006$ & $11.26_{-0.28}^{+0.02}$  & $1.03 \pm 0.12$ & $5.0 \pm 0.7$ \\
240892 & 150.6432950 & 2.0073169 & $1.5494\pm 0.0009$ & $11.29_{-0.13}^{+0.19}$  & $1.29 \pm 0.13$ & $3.0 \pm 0.3$ \\
205612 & 150.6542714 & 1.9233323 & $1.6751\pm 0.0045$ & $11.15_{-0.06}^{+0.13}$  & $2.08 \pm 0.28$ & $2.4 \pm 0.3$ \\
251833 & 150.6293675 & 2.0336620 & $1.4258\pm 0.0006$ & $10.99_{-0.05}^{+0.27}$  & $1.61 \pm 0.09$ & $2.1 \pm 0.1$ \\
228121 & 150.5936156 & 1.9754018 & $1.8084\pm 0.0015$ & $11.39_{-0.03}^{+0.14}$  & $2.78 \pm 0.86$ & $4.1 \pm 0.9$ \\
321998 & 150.7093826 & 2.1863891 & $1.5226\pm 0.0009$ & $11.28_{-0.09}^{+0.12}$  & $1.83 \pm 0.34$ & $4.4 \pm 0.6$ \\
299038 & 150.7091894 & 2.1369001 & $1.8196\pm 0.0010$ & $11.30_{-0.15}^{+0.08}$  & $0.96 \pm 0.02$ & $1.9 \pm 0.0$ \\
519818 & 150.0124421 & 2.6406856 & $2.0879\pm 0.0010$ & $11.37_{-0.24}^{+0.28}$  & $1.69 \pm 0.36$ & $5.4 \pm 0.7$ \\
526785 & 150.0057599 & 2.6548908 & $1.2454\pm 0.0383$ & $11.44_{-0.30}^{+0.04}$  & $4.95 \pm 0.54$ & $3.1 \pm 0.1$ \\
528213 & 150.0199418 & 2.6592730 & $1.3950\pm 0.0004$ & $11.33_{-0.22}^{+0.32}$  & $3.06 \pm 0.27$ & $3.2 \pm 0.2$ \\
535544 & 150.0027420 & 2.6748955 & $1.2452\pm 0.0003$ & $11.68_{-0.30}^{+0.05}$  & $5.73 \pm 1.22$ & $4.2 \pm 0.5$ \\
531916 & 149.9828887 & 2.6689945 & $1.3569\pm 0.0333$ & $11.14_{-0.41}^{+0.28}$  & \nodata         & \nodata \\
533754 & 150.0190505 & 2.6731339 & $1.3956\pm 0.0003$ & $11.60_{-0.18}^{+0.04}$  & $2.42 \pm 0.08$ & $1.3 \pm 0.1$ \\
543256 & 150.0179738 & 2.6948240 & $1.4340\pm 0.0008$ & $11.29_{-0.22}^{+0.19}$  & $3.34 \pm 0.39$ & $3.7 \pm 0.3$ \\
401700 & 150.2914647 & 2.3715033 & $1.6501\pm 0.0003$ & $11.16_{-0.14}^{+0.25}$  & $1.73 \pm 0.40$ & $4.5 \pm 1.0$ \\
411647 & 150.2915999 & 2.3956325 & $1.6525\pm 0.0008$ & $11.56_{-0.31}^{+0.08}$  & $2.67 \pm 0.41$ & $1.8 \pm 0.3$ \\
406178 & 150.2881135 & 2.3813900 & $1.5718\pm 0.0037$ & $11.56_{-0.60}^{+0.16}$  & \nodata         & \nodata
\enddata
\tablenotetext{a}{The \texttt{GALFIT} run was carried out by fixing $n=1$ \citep[see][]{onodera:2012}.}
\tablecomments{
  Effective radii are circularized, i.e., $r_\textrm{e}=a\sqrt{q}$, 
  where $a$ and $q$ are semi-major axis length and axis ratio, respectively.
}
\end{deluxetable*}
\capstarttrue

\subsection{Sample}
We selected 24 quenched galaxies with robust spectroscopic redshifts in the range $1.25<z<2.09$, derived from spectra obtained with the MOIRCS instrument at the Subaru telescope. 
For a detailed description of the observations and data reduction, refer to \citet{onodera:2012}.  
Briefly, all objects were first selected as passive BzK galaxies \citep{daddi:2004:bzk} based on the near-IR-selected photometric catalog 
in the COSMOS field \citep{scoville:2007,mccracken:2010}, with two additional criteria: a photometric redshift $z\gtrsim1.4$ and no mid-IR detection at 24 $\mu$m
with MIPS at the \textit{Spitzer Space Telescope}.
The redshift selection allowed us to observe the 4000 \AA{} break in most cases with the adopted instrument setup. 
The objects were integrated for 7--9 hrs each, under $0.4$--$1.2$ arcsec seeing conditions,
with 0.7 arcsec slit width, and using the zJ500 grism, which gives a spectral resolution of $R \sim 500$. 
The data were reduced with the standard  \texttt{MCSMDP} package \citep{yoshikawa:2010}, 
including flat-fielding, bad pixel and cosmic-ray removal, sky subtraction, distortion correction, 
and wavelength calibration by OH sky lines. 
Flux calibration was perfomed with A0V stars taken in the same nights as the object spectra 
and the slit-loss was corrected by scaling the continuum flux to the \textit{J}- or \textit{H}-band broadband fluxes. 
We measured the redshifts by cross-correlating the spectra either with simple stellar population (SSP) template spectra \citep{bruzual:2003} 
or with stellar template spectra \citep{sanchezblazquez:2006:miles}. 

\begin{figure}
  \centering
  \includegraphics[width=\linewidth]{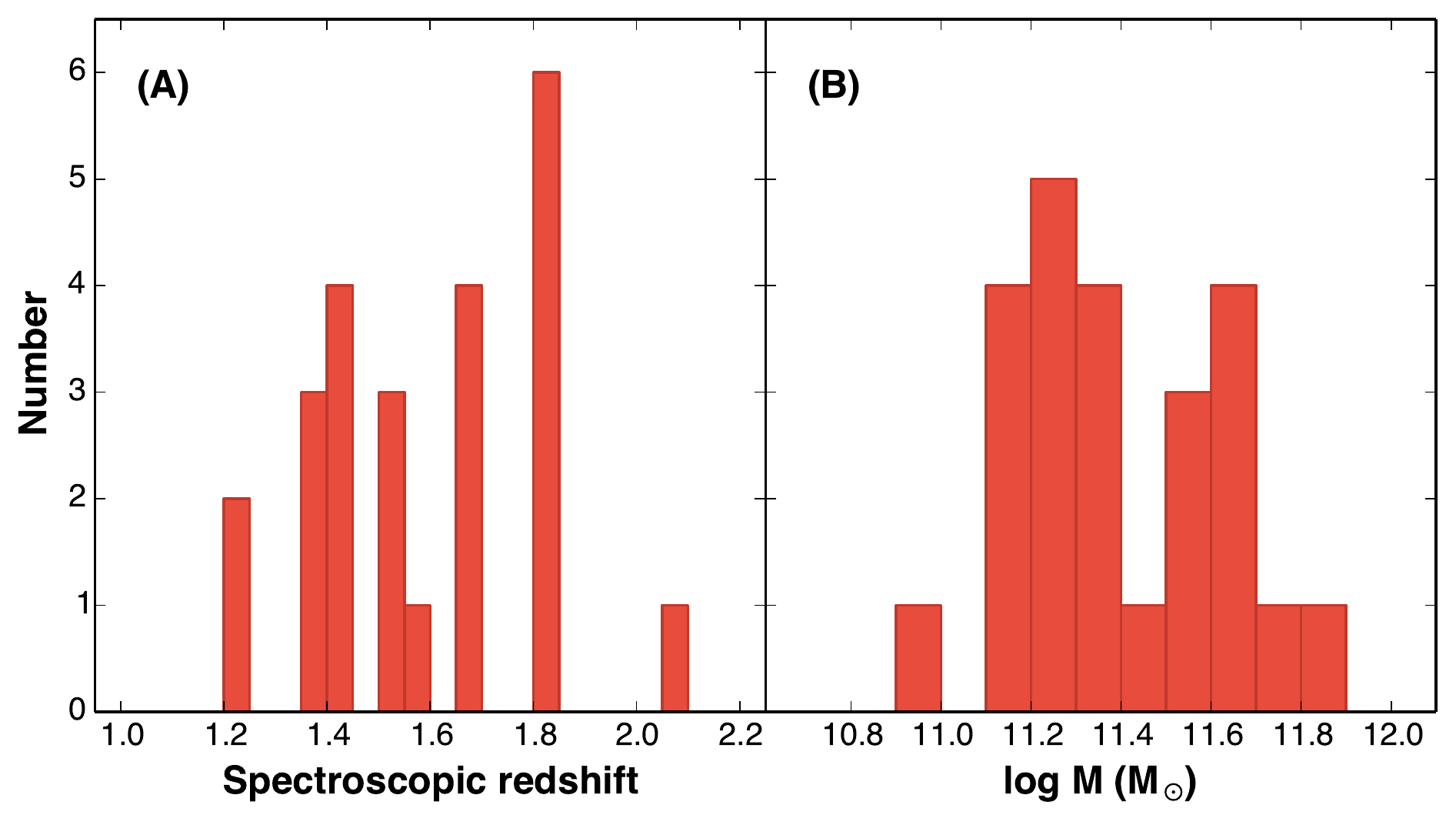}
  \caption{
    Histograms of (A) the spectroscopic redshifts and (B) the stellar masses of the 24 galaxies in the sample. 
  \label{fig:zhist}}
\end{figure}

\hyperref[fig:zhist]{Figure~\ref*{fig:zhist}A} shows the redshift distribution of the sample.
The mean and median redshifts are 1.56 and 1.59, respectively, with a standard deviation of 0.21.
Virtually all objects belong to redshift spikes.
Among the 24 objects, 14 have already been published \citep{onodera:2010:pbzk,onodera:2012} and the other 10 objects 
were identified during an observing run in 2011.  
We have removed 4 objects among the 18 identified in \citet{onodera:2012}: object ID 313880 was removed because of the presence of \textit{Spitzer}/MIPS 24\micron{} emission,
which may originate from star-forming activity or an AGN, 
and objects ID 209501, 253431, and 275414 were removed due to the lower confidence class for their spectroscopic redshifts. 

\subsection{Stellar masses and specific SFRs}
\label{sec:sedfitting}

We estimated stellar masses and SFRs by fitting synthetic spectral energy distributions (SEDs) to fluxes through broadbands extending from 
the  \textit{u} to the \textit{Spitzer}/IRAC 8.5 \micron{} band and 
using the \texttt{FAST} code\footnote{\url{http://astro.berkeley.edu/~mariska/FAST.html}} \citep{kriek:2009} 
as described in detail in \citet{onodera:2012}.  
In brief, we find a best-fit model within a grid of templates based on Charlot \& Bruzual (2007) stellar population synthesis models\footnote{\url{http://www.bruzual.org/cb07/}}
adopting an exponentially declining SFH with various $e$-folding times $\tau$. 
We allow for dust extinction adopting the  extinction law by \citet{calzetti:2000}.
The Salpeter initial mass function \citep[IMF;][]{salpeter:1955} was adopted, 
which may  be appropriate for galaxies as massive  as those in  our sample \citep{shetty:2014}.
Age, mass, $\tau$ and metallicity are then the results of the procedure, having fixed the redshifts  at the spectroscopic values. 
The resulting stellar masses\footnote{Sum of living stars and remnants.} are in the 
range $11<\log M / M_\odot < 11.9$,  with a median of $\log M/M_\odot = 11.4$ 
as shown in \hyperref[fig:zhist]{Figure~\ref*{fig:zhist}B}.

The MOIRCS spectra confirm the quenched nature of these galaxies, as they
show no emission lines, a strong break at rest-frame  4000 \AA{} and strong absorption lines typical of old stellar populations.
All objects have specific SFRs ($\text{sSFR} \equiv \text{SFR}/M$) below $10^{-11}\,\text{yr}^{-1}$, as derived from the best-fit exponentially declining SFH,  
which corresponds to a sSFR of about two orders of magnitude lower than that of main-sequence  galaxies at a similar redshift 
\citep[e.g.,][]{kashino:2013}.

\subsection{Structural parameters}
The effective radii, $r_\text{e}$, and \sersic indices, $n$, of the 10 additional objects from the 2011 observing run were measured 
in exactly the same way as was done for the rest of the sample presented in \citet{onodera:2012}.  
In brief, we used \texttt{GALFIT}\footnote{\url{http://users.obs.carnegiescience.edu/peng/work/galfit/galfit.html}}
version 3.0 \citep{peng:2002:galfit, peng:2010:galfit} to carry out a 2D \sersic profile fitting 
for the images taken in the F814W filter with the Advanced Camera for Surveys (ACS) on the  \textit{Hubble Space Telescope}
\citep[\textit{HST};][]{koekemoer:2007,massey:2010}.  
For each galaxy, we constructed the point spread function (PSF) from nearby unsaturated stars.  
We estimated the sky background using three different methods: 
(1) the sky as a free parameter; 
(2) the sky fixed to the so-called pedestal GALFIT estimate; 
and (3) the sky manually measured from empty regions near the sample galaxy.  
We used the midpoints between the minimum and maximum of these runs, and half of the ranges as associated errors, 
for effective radius, \sersic parameter, total magnitude, and axial ratio. 
\texttt{GALFIT} did not converge for objects 406178 and 531916 due to low surface brightness in the F814W filter. 

Object IDs, coordinates, spectroscopic redshift, stellar mass, effective radius, and \sersic index for all of the galaxies in  the sample are listed in \autoref{tab:objprop}.

\subsection{Composite spectrum}

\begin{figure*}
  \centering
  \includegraphics[width=0.9\linewidth]{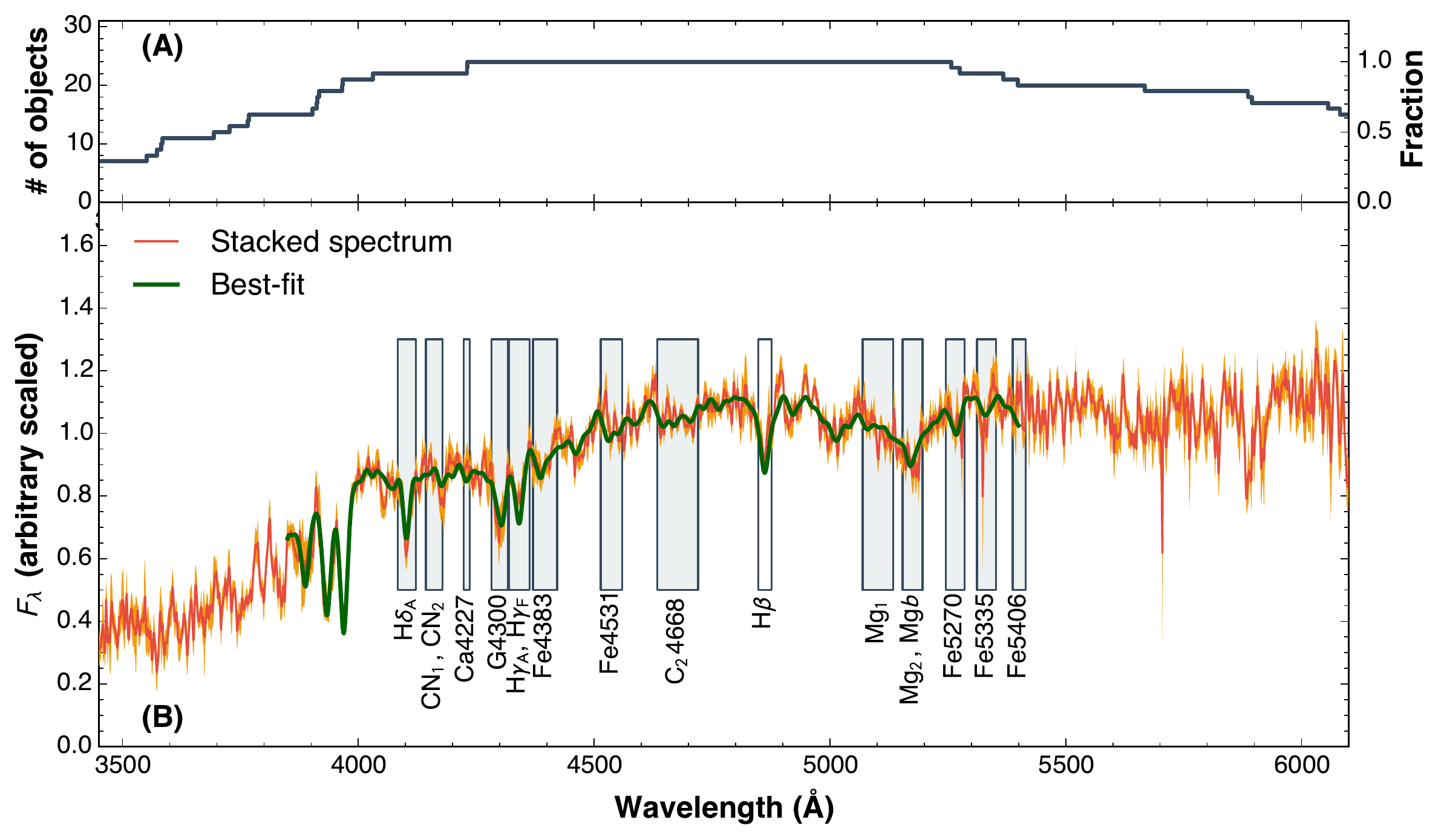}
  \caption{
    Composite rest-frame optical spectrum of the 24 quenched galaxies at $z\sim1.6$.
    (A): the number (left axis) and fraction (right axis) of spectra that have been  stacked at each wavelength. 
    (B): the stacked spectrum and associated $1\sigma$ error (orange solid line and filled region, respectively).  
    The green solid line shows the best-fit combination of stellar spectra \citep{sanchezblazquez:2006:miles}, 
    using \ppxf \citep{cappellari:2004:ppxf}. 
  The rectangles show the wavelength ranges used to measure the Lick indices of the stacked spectrum, though the H$\beta$ index is not used  in our stellar population analysis. 
    \label{fig:stackspec}}
\end{figure*}

In order to achieve the high S/N needed for 
the measurement of the Lick indices, 
we constructed a composite spectrum by stacking the individual galaxy spectra
in the same way as described in \citet{onodera:2012}: 
all spectra were registered to the rest-frame wavelength by linearly interpolating on a 1 \AA{} interval, 
and normalized by the mean flux at $4500<\lambda(\text{\AA})<5200$ in the rest-frame. 
Noise spectra were also registered to the same rest-frame wavelength scale, 
but interpolated in quadrature. 
Then the spectra were co-added with weights proportional to the inverse square of 
the S/N at each wavelength pixel. 

To correct for possible sampling bias, we computed the associated noise spectrum with the \textit{jackknife} method as
\begin{equation}
  \sigma_{\text{Jack}}^2 = \frac{N-1}{N}\sum_{i=1}^{N}(f-f_{(i)})^2,
  \label{eq:jk1}
\end{equation}
where $N\; (=24)$ is the number of objects in the sample, 
$f$ is the flux of the stacked spectrum of all $N$ spectra, 
and $f_{(i)}$ is the flux of a stacked spectrum made of $N-1$ spectra 
by removing the $i$th spectrum.  
We then corrected the stacked spectrum for the bias using 
\begin{equation}
  f'=f-(N-1)(\langle f_{(i)} \rangle -f).
  \label{eq:jk2}
\end{equation}
The typical correction factor is very small ($<1\%$) 
in the wavelength range studied in this work, 
but it is a more realistic estimate of noise due to OH sky line residuals
which vary significantly from  pixel to pixel. 
We have adopted $f'$ and $\sigma_\text{Jack}$ as the final stacked spectrum and $1\sigma$ noise spectrum, respectively.  

The resulting composite spectrum, corresponding to an equivalent integration time of about 200 hr on Subaru/MOIRCS, 
is shown in \autoref{fig:stackspec}.  
This figure shows several clear absorption features,
including the 4000~\AA{} break, 
\ion{Ca}{2} H+K lines, 
various Balmer absorption lines, 
the \textit{G}-band,
and Mg and Fe absorption features,
which will be used in the next section to derive the stellar population parameters.

%%%%%%%%%%%%%%%%%%%%%%%%%%%%%%%%%%%%%%%%%%%%%%%%%%%%%%%%%%%

\section{Measurement of Lick absorption line indices}
\label{sec:lick_measurement}

We measured the Lick indices on the stacked spectrum using 
\texttt{LECTOR}\footnote{\url{http://www.iac.es/galeria/vazdekis/vazdekis\_software.html}}, 
following the definition given by \citet{worthey:1997} and \citet{trager:1998}. 
We used the \textit{jackknife} method to measure each index following
\hyperref[eq:jk1]{Equation~\ref*{eq:jk1}} and \hyperref[eq:jk2]{Equation~\ref*{eq:jk2}}: 
we measured the indices on each $f_{(i)}$ and derived the bias-corrected index values and
corresponding $1\sigma$ error bars. 
The Lick indices measured on the stacked spectrum with the observed spectral resolution 
are listed in the second column in \autoref{tab:lick}. 

\capstartfalse
\begin{deluxetable}{ccc}
  \tablewidth{0pt}
  \tablecolumns{11}
  \tablecaption{Lick indices measured in the stacked spectrum \label{tab:lick}}
  \tablehead{
    \colhead{Index} &
    \colhead{Raw Measurement} &
    \colhead{Broadening Corrected}
  }
  \startdata
  \hda    & $ \phantom{-}4.27 \pm 0.69$  &  $ \phantom{-}4.26 \pm 0.72$ \\
  \hdf    & $ \phantom{-}3.26 \pm 0.74$  &  $ \phantom{-}3.61 \pm 0.79$ \\
  \cnone  & $-0.08 \pm 0.02$  &  $-0.08 \pm 0.02$ \\
  \cntwo  & $-0.03 \pm 0.03$  &  $-0.02 \pm 0.03$ \\
  Ca4227  & $-0.15 \pm 0.48$  &  $-0.23 \pm 0.75$ \\
  \gband  & $ \phantom{-}4.45 \pm 1.31$  &  $ \phantom{-}4.90 \pm 1.34$ \\
  \hga    & $-0.88 \pm 1.34$  &  $-0.95 \pm 1.38$ \\
  \hgf    & $ \phantom{-}1.14 \pm 0.73$  &  $ \phantom{-}1.38 \pm 0.80$ \\
  Fe383   & $ \phantom{-}2.67 \pm 1.38$  &  $ \phantom{-}3.57 \pm 1.59$ \\
  Fe4531  & $ \phantom{-}0.51 \pm 1.12$  &  $ \phantom{-}0.70 \pm 1.38$ \\
  \ctwo   & $ \phantom{-}4.71 \pm 1.09$  &  $ \phantom{-}5.51 \pm 1.25$ \\
  \hbeta  & $ \phantom{-}2.36 \pm 0.60$  &  $ \phantom{-}2.68 \pm 0.73$ \\
  \mgone  & $ \phantom{-}0.09 \pm 0.02$  &  $ \phantom{-}0.10 \pm 0.02$ \\
  \mgtwo  & $ \phantom{-}0.18 \pm 0.01$  &  $ \phantom{-}0.18 \pm 0.01$ \\
  \mgb    & $ \phantom{-}2.13 \pm 0.72$  &  $ \phantom{-}2.88 \pm 0.97$ \\
  Fe5270  & $ \phantom{-}1.53 \pm 0.89$  &  $ \phantom{-}2.09 \pm 1.18$ \\
  Fe5335  & $ \phantom{-}1.69 \pm 0.71$  &  $ \phantom{-}3.25 \pm 1.38$ \\
  Fe5406  & $ \phantom{-}0.96 \pm 1.05$  &  $ \phantom{-}1.88 \pm 2.03$
\enddata
\tablecomments{
  \cnone, \cntwo, \mgone, and \mgtwo{} indices are in a unit of magnitude;
  otherwise, indices are in a unit of \AA{}.
  \hdf{} and \hbeta{} indices are not used in the analysis presented in this work. 
}
\end{deluxetable}
\capstarttrue

\subsection{Broadening measurement and stellar velocity dispersion}

Measurements of Lick indices strongly depend on the broadening of the spectra:
the larger the broadening, the shallower the absorption features
and the more likely that part of the spectral feature falls outside the index windows. 
The total broadening ($\sigma_\text{tot}$) is a combination of the intrinsic stellar velocity dispersion ($\sigma_\star$) of the object, 
instrumental resolution ($\sigma_\text{instr}$), and stacking procedure reflecting redshift errors ($\sigma_\text{stack}$), 
and can be expressed as $\sigma_\text{tot}^2=\sigma_\star^2+\sigma_\text{instr}^2+\sigma_\text{stack}^2$ \citep{cappellari:2009:gmass}.  
We measured $\sigma_\text{tot}=457\pm23\kms$ by using the penalized pixel-fitting method 
\citep[\ppxf\footnote{\url{http://purl.org/cappellari/software}};][]{cappellari:2004:ppxf} with template stellar spectra from the MILES database\footnote{\url{http://miles.iac.es/}} \citep{sanchezblazquez:2006:miles}.
During the \ppxf run, we masked wavelength ranges that are potentially contaminated by emission lines,
namely H$\gamma$, H$\beta$, and [\ion{O}{3}]$\lambda\lambda4959,5007$. 
%% despite the fact that contamination from these emission lines appears to be always below $1\sigma$ of the noise spectra. 
The best-fit combination of the templates is shown with the green solid line in \autoref{fig:stackspec}. 

Adopting the same procedure as \citet{cappellari:2009:gmass}, 
$\sigma_\text{stack}$ can be computed using $\sigma_\text{stack}\approx c \Delta z /(1+z)$, 
where $c$ is the speed of light and $\Delta z$ is the redshift error. 
We computed $\sigma_\text{stack}$ for all objects used for stacking, 
and derived the average as $\sigma_\text{stack}=101 \pm 83$ \kms{} 
by using the bi-weight estimator \citep{beers:1990}. 
This $\sigma_\text{stack}$ may be a conservative estimate, 
because objects with larger redshift errors have lower S/Ns and thus have less weight in the stacked spectrum. 
Adopting the instrumental resolution of $\sigma_\text{instr}=300 \pm 7$ \kms{} \citep{onodera:2010:pbzk}, 
by combining these numbers together, the stellar velocity dispersion of the stacked spectrum 
is estimated to be $\sigma_\star=330 \pm 41$ \kms{}. 
This is in good agreement with the average stellar velocity dispersion, $\langle \sigma_\text{inf} \rangle = 298 \pm 71$ \kms{}, 
inferred from the structural parameters and stellar masses of the objects \citep{bezanson:2011}.

\subsection{Broadening correction}

The resulting $\sigma_\text{tot}$ is in general larger than that of the original Lick/IDS system which is $\simeq200$--$360$ \kms{} 
depending on the indices. 
Therefore, the indices measured on our stacked spectrum have to be corrected to the Lick resolution.  
For this purpose, we computed the Lick indices on the MILES SSP models \citep{vazdekis:2010} 
by convolving the model spectra with the original Lick resolution \citep{worthey:1997,schiavon:2007} 
and our resolution of $\sigma_\text{tot}=460 \kms$. 
The MILES SSP models\footnote{\url{http://miles.iac.es/pages/webtools/tune-ssp-models.php}} 
were constructed with a unimodal IMF with a slope of 1.3 (equivalent to the Salpeter IMF), 
$-2.32<[\text{M/H}]<0.22$ and age from 0.06 Gyr to that of the Universe at the corresponding redshift.  
We restricted the models to be in the safe range of parameters by setting  \texttt{Mode=SAFE}. 
Comparing the indices measured with various broadenings, 
we found tight linear relations ($\lesssim 1\%{}$ scatter) for all the indices 
and we used these relations to convert our values 
to those with the Lick/IDS resolution. 
The Lick indices corrected for the spectral resolution are listed in the third column in \autoref{tab:lick}.

%%%%%%%%%%%%%%%%%%%%%%%%%%%%%%%%%%%%%%%%%%%%%%%%%%%%%%%%%%%

\section{Stellar population parameters from Lick indices}
\label{sec:stellarpop}

\begin{figure*}
  \centering
  \includegraphics[width=0.95\linewidth]{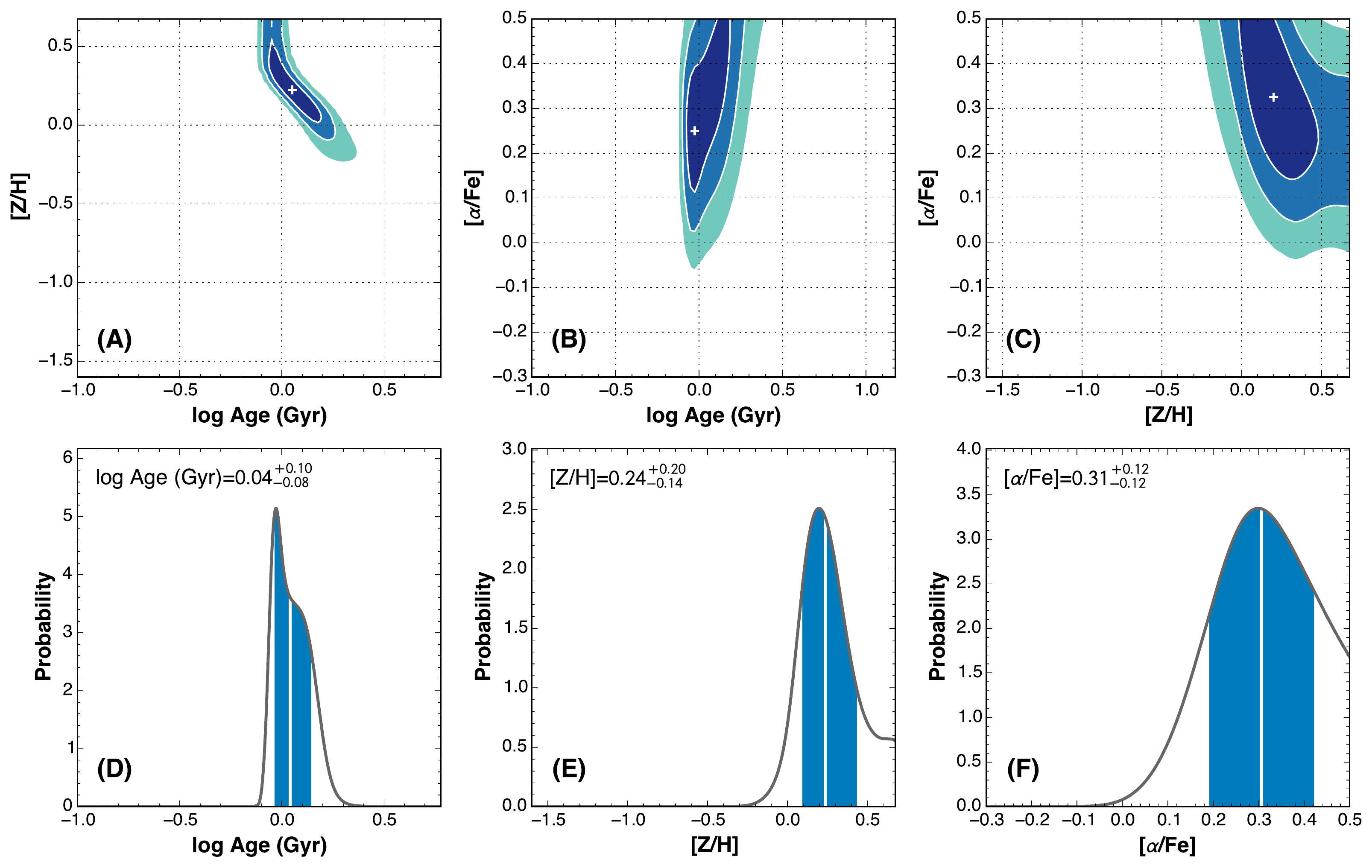}
  \caption{
    The probability distributions of stellar population parameters for the quenched galaxies at $z\sim1.6$. 
    Top: two-dimensional probability distributions in the 
    (A) age--\zoh, (B) age--\afe, and (C) \zoh--\afe planes. 
    Cross symbols show the peak of the probability distributions and the contours indicate 
    the areas that enclose 68.2\%{}, 95.4\%{}, and 99.7\%{} of the probability distributions, respectively. 
    Bottom: one-dimensional probability distributions for (D) age, (E) \zoh, and (F) \afe. 
    The vertical white solid lines indicate the median values, 
    and the left and right edges of the filled regions define the 16 and 84 percentiles, respectively. 
    \label{fig:prob_all}}
\end{figure*}

In the following analysis we use a series  of Lick indices for which synthetic models accurately reproduce those of Galactic globular clusters,
as shown by \citet{thomas:2011:tmj},  namely: 
\hda, \hga, \hgf, \cnone, \cntwo, Ca4227, \gband, \ctwo, \mgone, \mgtwo, \mgb, Fe4383, Fe4531, Fe5270, Fe5335, and Fe5406.
The \hbeta{} index is widely used as an age indicator since it is less contaminated by 
metal lines, but it was not used in this analysis for the following two reasons.
First, there may be a contribution from emission partially filling the absorption.
Comparing the stacked spectra and best-fit template returned by pPXF,
we measured a $3\sigma$ upper limit of $\simeq 1$ \AA{} to the equivalent width
of the \hbeta{} emission line.
On the other hand, the higher-order Balmer lines such as H$\gamma$ and H$\delta$ are less affected by 
emission, with an equivalent width $\lesssim 0.5$ \AA.
Second, the \hbeta{} indices in Galactic globular clusters are too weak for 
even the oldest SSP models to reproduce \citep{poole:2010}.
We would like to note, however, that the inclusion of \hbeta{} to the 
set of indices above does not affect the results and conclusions presented here.

We derived the stellar population parameters, age, \zoh, and \afe, 
by comparing the Lick indices measured 
on the composite spectrum with those calculated from SSP 
models with variable abundance ratios\footnote{\url{http://www.icg.port.ac.uk/~thomasd/tmj.html}} \citep{thomas:2011:tmj}. 
Since the models give absolute fluxes by making use of 
flux-calibrated stellar spectral templates, the observed indices 
do not need to be converted to the Lick/IDS response. 
The models span the parameter space   $0.1<\text{age/Gyr}<15$, $-2.25<\zoh<0.67$, and $-0.3<\afe<0.5$ and adopt the Salpeter IMF. 

For each index, we constructed a three-dimensional model grid of \logage, \zoh, and \afe 
with an uniform interval of 0.025 dex for all three parameters
and on this grid we computed the 
$\chi^2 = \sum_{\text{index}} {\left(I_{\text{index}}^{\text{obs}}-I_{\text{index}}^{\text{model}}\right)^2}/{\sigma_{\text{index}}^2}$, 
where $I_{\text{index}}^{\text{obs}}$ and $I_{\text{index}}^{\text{model}}$ are observed and model indices, respectively, 
and $\sigma_{\text{index}}$ is the $1\sigma$ error of the observed indices. 
Then we converted the $\chi^2$ to a probability distribution $p\propto\exp\left(-\chi^2/2\right)$, 
and normalized it such that $\int p\,\text{d}\logage\,\text{d}\zoh\,\text{d}\afe = 1$. 

The resulting best-fit stellar population parameters are shown as white plus signs in \autoref{fig:prob_all},
which corresponds to the minimum reduced-$\chi^2$ of $\sim 2.0$.
The reduced-$\chi^2$ drops to 1.37 by removing the CN indices from the fit,
while the best-fit parameters do not change at all.
This poor fit due to the CN indices is not surprising,
as nitrogen is not a free parameter in the fit \citep[see also][]{thomas:2011:tjm,thomas:2011:tmj}. %%Thomas et al. (2011).
Confidence regions enclosing 68.2\%, 95.4\%, and 99.7\%{} of the total volume of the projected probability distributions 
are shown in the top panels of \autoref{fig:prob_all}.
The probability distributions are marginalized to a one-dimensional parameter space 
(bottom panels in \autoref{fig:prob_all}), 
by adopting median values for each stellar population parameter 
and 16 and 84 percentiles as corresponding 1$\sigma$ confidence intervals, respectively.
In this way we have obtained $\log(\text{age}/\text{Gyr})=0.04_{-0.08}^{+0.10}$, 
$\zoh=0.24_{-0.14}^{+0.20}$, and $\afe=0.31_{-0.12}^{+0.12}$ for the composite stellar population of our 24 galaxies.
Of course, these uncertainties  only reflect random errors associated with the measurement of the various indices and the quality of the fits, whereas 
systematic errors coming from the  synthetic models may be significantly larger \citep[e.g.,][]{trager:2008}.

%%%%%%%%%%%%%%%%%%%%%%%%%%%%%%%%%%%%%%%%%%%%%%%%%%%%%%%%%%%

\section{Discussion}
\label{sec:discussion}

\subsection{Bias in the sample}
\label{sec:uvj}
Spectroscopically identified samples could be biased toward relatively bluer, lower mass-to-light ratio objects,
especially at $z>1.4$ as pointed out by \citet{vandesande:2014}. 
Although our sample is nearly complete down to $K_\text{AB}\simeq21.5$ \citep{onodera:2012},
in this section we investigate potential biases that may affect our sample.

\begin{figure}
  \centering
  \includegraphics[width=0.9\linewidth]{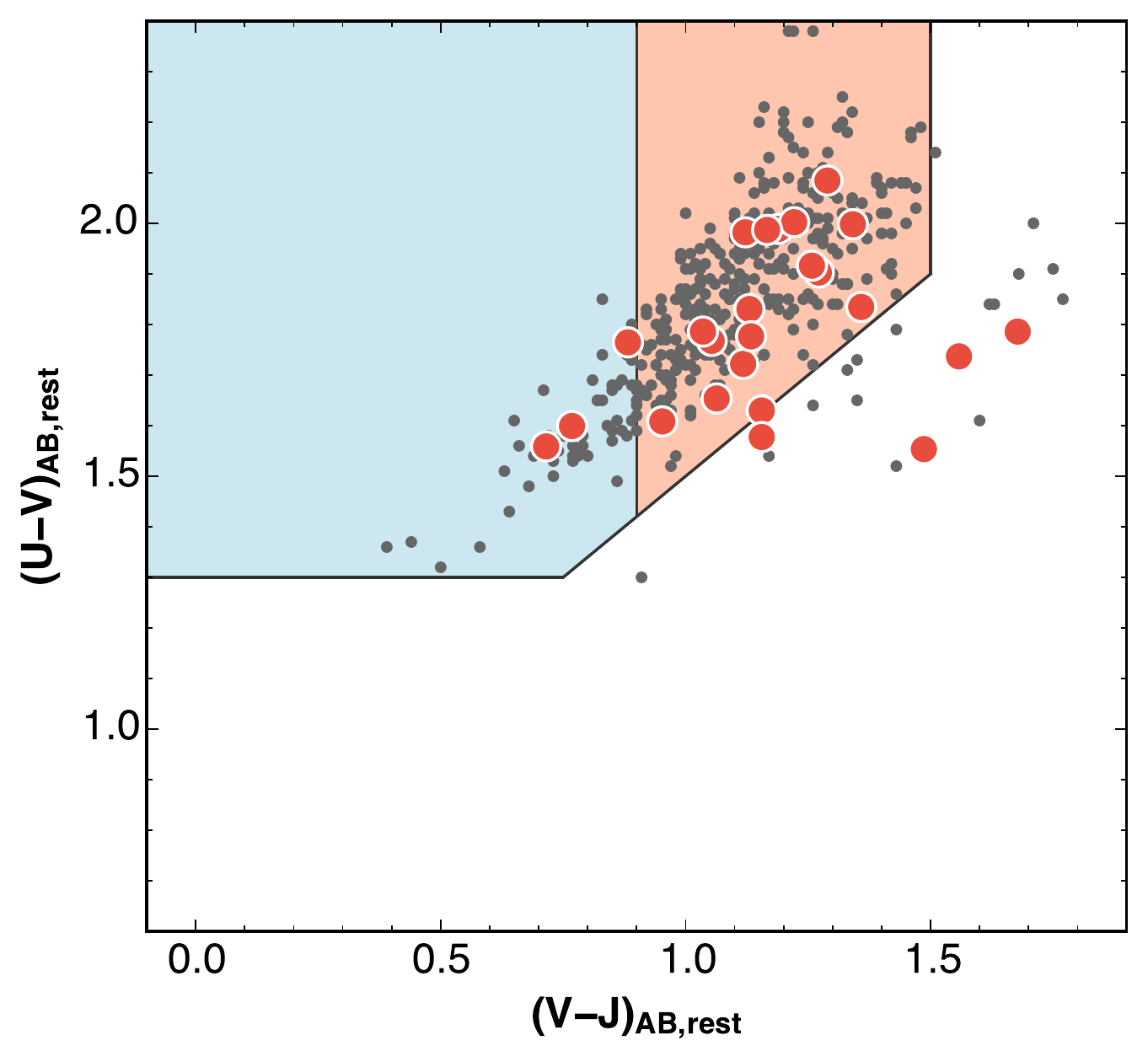}
  \caption{
      Rest-frame \textit{UVJ} color--color diagram.
      Red circles are our quenched galaxies at $\mz=1.6$ with \textit{UVJ} colors
      computed as described in \hyperref[sec:sedfitting]{Section~\ref*{sec:uvj}}.
      Gray dots represent photometrically selected passive BzK galaxies \citep{daddi:2004:bzk}
      with reliable photometry, non-detection in \textit{Spitzer}/MIPS 24 $\mu$m, and $K_\text{AB}<21.5$
      taken from \citet{muzzin:2013:catalog}.
      Blue and red areas indicate regions to separate young and old quiescent galaxies, respectively, 
      while rest of the diagram classifies galaxies as star-forming \citep{whitaker:2012}. 
      \label{fig:uvj}}
\end{figure}

We computed rest-frame $\mathit{U}-\mathit{V}$ and $\mathit{V}-\mathit{J}$ colors of our sample 
and compared them with the photometric sample of passive BzK-selected galaxies in \autoref{fig:uvj}.
Since we have extracted the photometric sample for this comparison from a public catalog
in the COSMOS field by \citet{muzzin:2013:catalog}, %% as described below,
we obtained the rest-frame colors using the \citeauthor{muzzin:2013:catalog} photometry. 
We ran \texttt{EAZY}\footnote{\url{http://www.astro.yale.edu/eazy/}} \citep{brammer:2008}
using the identical template set and input parameters as those used in \citeauthor{muzzin:2013:catalog}, 
except that redshifts are fixed to the spectroscopic redshifts, 
and computed the rest-frame colors for \textit{U}- and \textit{V}-bands of \citet{maizapellaniz:2006}
and 2MASS \textit{J}-band.
For the photometric sample extracted from the \citeauthor{muzzin:2013:catalog} catalog,
we required the photometry flag of unity, non-detection in \textit{Spitzer}/MIPS 24 $\mu$m,
and $K_\text{AB}<21.5$ to match the selection of our sample,
in addition to the standard passive BzK selection criteria \citep{daddi:2004:bzk}. 

\autoref{fig:uvj} shows that the majority of
our sample are classified as ``old quenched'' according to the \citet{whitaker:2012} criterion
and the overall distribution appears to be consistent with that of the parent sample. 
Quantitatively, the median colors of our sample 
and the parent sample are $(\mathit{V}-\mathit{J}, \mathit{U}-\mathit{V})=(1.16, 1.78)$
and $(\mathit{V}-\mathit{J}, \mathit{U}-\mathit{V})=(1.10, 1.85)$, respectively.
Note that the typical uncertainty in the rest-frame colors is $0.1$--$0.2$ mag \citep[e.g.,][]{williams:2009}
and the above differences are consistent within it.
One might argue that the \textit{UVJ} selection cannot perfectly isolate
the passive BzK population in the quiescent section of the diagram or vice versa,
even though there appears to be no signature of emission lines in our spectroscopic sample. 
However, we would like to note that both \textit{UVJ} and BzK
diagnostics select quenched galaxies only in a statistical sense 
and some degree of contamination outside of the quiescent region is not entirely unexpected
(see \citealt{moresco:2013} for a detailed comparison of various selection methods of passive galaxies at $z<1$).

\citet{whitaker:2013} have derived stellar population ages of \textit{UVJ}-selected quiescent galaxies at $1.4<z<2.2$
identified in \textit{HST}/WFC3 G141 grism observations from the 3D-HST survey \citep{brammer:2012}.
The average age is $1.3_{-0.3}^{+0.1}$ Gyr for all quiescent galaxies,
which is in excellent agreement with that of our quenched galaxies at similar redshift.
They also split the sample into \textit{young} and \textit{old} quiescent galaxies,
and the derived ages for both classes are consistent with ours within the uncertainties.

Comparisons of the rest-frame colors with the parent passive BzK sample
and of the stellar population ages with the galaxies with similar colors from \citet{whitaker:2013}
suggest that our spectroscopic sample is not likely to be biased toward younger ages
and appears to be representative of the quenched galaxy population at $\mz=1.6$.
However, the comparison discussed here is based on the \textit{K}-selected sample. 
Therefore, the lack of apparent bias in our spectroscopic sample with respect to the photometric one
does not necessarily mean that there is no bias compared to the entire (e.g., mass-selected) quiescent population.

\subsection{Redshift evolution of stellar populations in quenched galaxies}

\begin{figure*}
  \centering
  \includegraphics[width=0.7\linewidth]{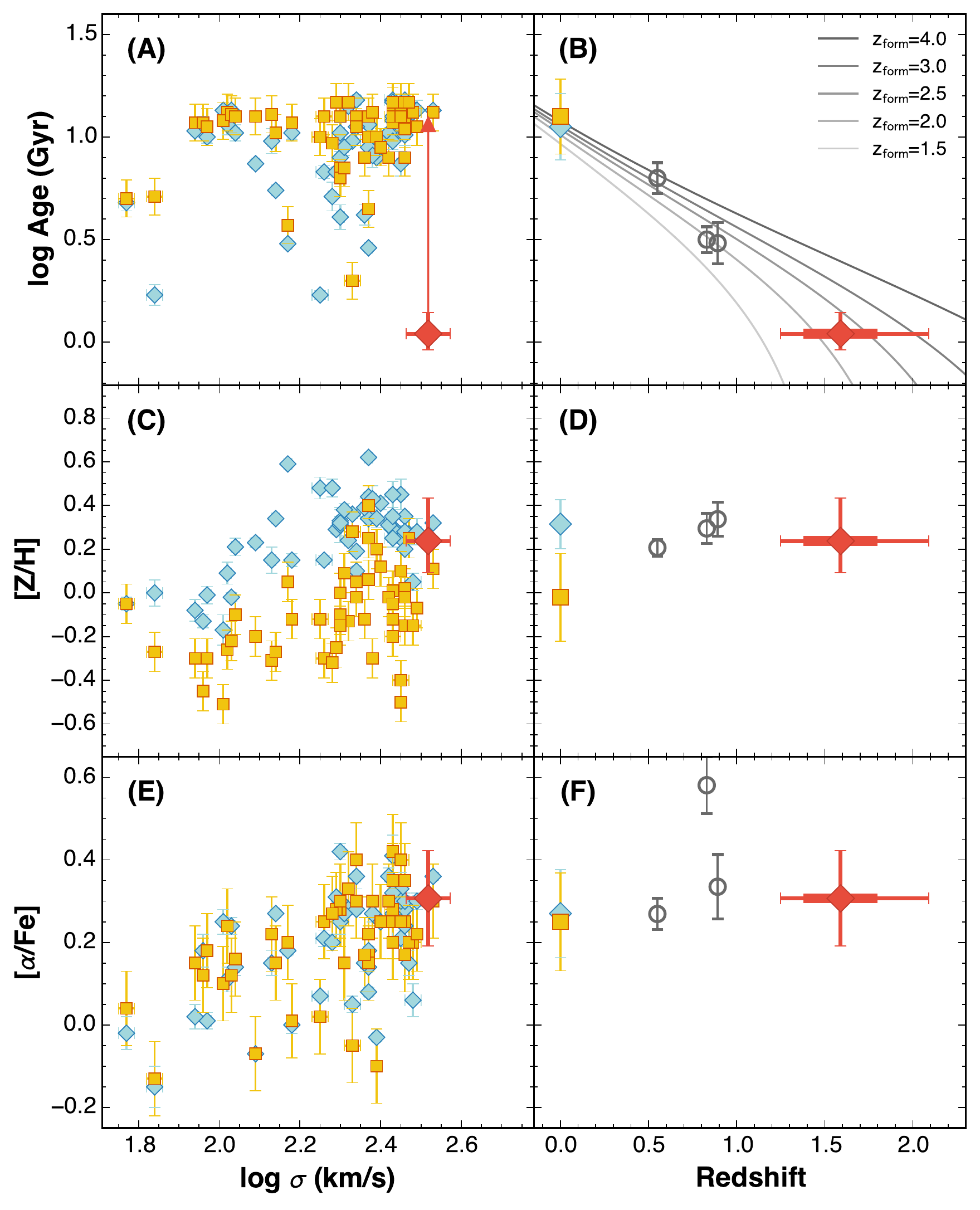}
  \caption{
    Stellar population parameters as a function of stellar velocity dispersion (A), (C), (E) and redshift (B), (D), (F). 
    Shown are the luminosity-weighted age (A), (B), \zoh (C), (D), and the \afe ratio (E), (F) of stellar populations in quenched galaxies at various redshifts. 
    In each panel, the red symbol represents our measurement at $\mz=1.6$. 
    In the right panels, the thick and thin error bars correspond to the standard deviation 
    and range of redshift of the sample, respectively. 
    Blue and orange symbols show $z=0$ values within $r_\text{e}/8$ and $r_\text{e}$, respectively, 
    of local quenched galaxies \citep{spolaor:2010}.
    In the right panels, blue and orange points are the corresponding median properties of the local sample with $\sigma>200$ \kms{}
    with the $1\sigma$ scatter of the corresponding distribution. 
    The red arrowhead in panel (A) shows the ending point of a purely passive evolution of 
    the stellar populations of our sample galaxies down to $z=0$. 
    Gray circles in the right panels show the values for massive quenched galaxies with $\log\sigma(\kms)>2.24$ 
    in intermediate redshift clusters \citep{jorgensen:2013}. 
    Their measurements were aperture corrected 
    in order to match the nuclear measurements at $z=0$. 
    In panel (B), the gray solid lines show, from thin to thick, 
    the age of simple stellar populations made at a formation redshifts  from 1.5 to 4.0, as indicated in the insert. 
    \label{fig:pop_sigmaz}}
\end{figure*}

\autoref{fig:pop_sigmaz} compares the stellar population luminosity-weighted age, 
\zoh, and \afe of various sets of quenched galaxies at different redshifts. 
Having derived an age 
$\log(\text{age/Gyr})=0.04_{-0.08}^{+0.10}$, or $\simeq1.1$ Gyr, from the composite spectrum,
we calculate that the passive aging over $\sim 9.5$ Gyr from $\mz=1.6$ to $z=0$, i.e., to an age of $\sim 11$ Gyr, will bring
the stellar populations of these galaxies to the point indicated by
the arrowhead in \hyperref[fig:pop_sigmaz]{Figure~\ref*{fig:pop_sigmaz}A}. 
This is in excellent agreement with the
luminosity-weighted ages of stellar populations in local quenched
galaxies of similar velocity dispersions by \citet{spolaor:2010}, 
who derive the stellar population parameters via Lick indices using stellar population models
from \citet{thomas:2003:tmb}, 
which are equivalent in this respect to the  \citet{thomas:2011:tmj} models adopted in this study.

\hyperref[fig:pop_sigmaz]{Figure~\ref*{fig:pop_sigmaz}B} shows the age--redshift
relations for different values of the formation redshift $z_\text{f}$
(gray solid lines), assuming a single burst of star formation at $z_\text{f}$.  
In this SSP approximation,
the average age of $\sim 1.1$ Gyr indicates a formation redshift of
$2\lesssim z_\text{f} \lesssim 2.5$ for the bulk of the stars.
\hyperref[fig:pop_sigmaz]{Figure~\ref*{fig:pop_sigmaz}B} 
also shows the stellar population ages of
similarly massive, quenched galaxies in clusters at $z=0.54$, $0.83$,
and $0.89$ \citep{jorgensen:2013}.
These are currently the only ages for quenched galaxies at these intermediate epochs with comparable $\sigma$
that have been derived in a homogeneous way similar to our galaxies, i.e., 
using the Lick indices and the \citet{thomas:2011:tmj} models to derive the stellar population parameters. 
Apart from a possible (small) bias due to the special location in high density peaks, 
these lower redshift cluster galaxies fit remarkably  well  along
the evolutionary trend expected for pure passive evolution from 
$\mz=1.6$ to $z=0$.

Furthermore, \autoref{fig:pop_sigmaz} shows that the derived metallicity $\zoh=0.24_{-0.14}^{+0.20}$,
or $\simeq 1.7\,\text{Z}_\odot$,
of our composite $\mz=1.6$ galaxies is the same as that of  
local quenched galaxies of similar velocity dispersion \citep{spolaor:2010}.
The two different colors---blue and orange---in the figure correspond to
the values measured at the central part of galaxies (within $r_\text{e}$/8) and within the effective radius, respectively. 
The aperture used to extract the spectra of the $\mz=1.6$ sample is about $(2$--$3) r_\text{e}$.
In the case that the galaxy is well resolved,
stellar population parameters derived within large radii (e.g., $r>r_\text{e}$)
converge to those integrated to the effective radius \citep{kobayashi:1999}.
However, our sample was observed with a FWHM of the PSF corresponding to $\gtrsim 2r_\text{e}$. 
\citet{vandesande:2013} have examined the aperture correction for velocity dispersion measurements
by taking PSF and aperture effects into account for a \sersic profile and
we have followed the same approach to investigate these effects on the derived metallicity.
Since radial metallicity gradients in the local quenched galaxies are very steep
\citep[$\Delta \zoh / \Delta \log r = -0.23$; ][see also \citealt{kuntschner:2010,koleva:2011}]{spolaor:2010},
the change in \zoh integrated within $r_\text{e}/8$ and out to $r \gg r_\text{e}$
is found to be only $\lesssim 0.05$ dex
when coupled with a \sersic profile with $n\gtrsim1$ and large PSF size compared to the effective radius. 
Therefore, our measured \zoh appears to be more representative of the central value. 
Note that the measured indices, hence the derived age, \zoh, and \afe presented in
\citet{jorgensen:2013} for intermediate-$z$ cluster galaxies
have already been corrected to the values within 3.4 arcsec at the distance of the Coma cluster,
based on measurements of index gradients of local galaxies \citep[][see also \citealt{jorgensen:1995,jorgensen:2005}]{jorgensen:2013}. 

The $\alpha$-to-iron abundance ratio measured for the $\mz=1.6$ quenched galaxies, $\afe=0.31_{-0.12}^{+0.12}$,
or $\sim 2$ times the solar value, lies precisely on the $z=0$ \afe--$\sigma$ relation 
\citep[\autoref{fig:pop_sigmaz}; e.g.,][]{kuntschner:2010,spolaor:2010,thomas:2010}, 
indicating little or no evolution at a given velocity dispersion of the \afe ratio over the past $\sim 10$ Gyr. 

Thus, it appears that the chemical composition is indeed \textit{frozen in} during passive evolution from $\mz=1.6$ to 0. 
We conclude that the stellar population content, i.e., their age, metallicity and $\alpha$-element enhancement 
of our $\mz=1.6$ galaxies, qualifies them as possible progenitors of similarly massive quenched galaxies at $z=0$, 
from purely passive evolution.

\subsection{SFHs of quenched galaxies at $\mz=1.6$ and their precursors}

We turn now to the other side in cosmic times, i.e., toward higher redshifts and earlier epochs,
trying to identify possible progenitors of our quenched galaxies at  $\mz=1.6$. 
At higher redshift  star-forming galaxies dominate 
the massive galaxy population \citep[e.g.,][]{ilbert:2013,muzzin:2013:mf} and it is among them that precursors should be sought.  
The average age $\sim 1.1$ Gyr of our quenched galaxies at $\mz=1.6$ indicates a formation redshift of $z_\text{f}\simeq2.3$ for their stellar populations, 
as also indicated in \hyperref[fig:pop_sigmaz]{Figure~\ref*{fig:pop_sigmaz}B}.
As already mentioned, having been derived from fits to SSPs, 
this ``age'' is essentially a measure of the time elapsed since their star formation was quenched. 
Indeed, if the SFR was rapidly increasing, % prior to the quenching, 
as expected for the evolution of main-sequence galaxies at such high redshift \citep{renzini:2009,peng:2010}, 
then the corresponding formation redshift must also be close to the quenching epoch,
as most of the stars formed just prior to the quenching.
By the same token, these authors argue that an $\alpha$-element enhancement relative to the solar abundance ratio 
is the natural outcome of such a (quasi-exponential) rapid increase in SFR. 

Following \citet{thomas:2005}, we use the measured \afe ratio
to estimate the star formation timescale of the precursors as 
$\log (\Delta t/\mathrm{Gyr}) \approx 6\times\left(\frac{1}{5} -\afe\right) = -0.66 \pm 0.72$.
Given the large error bars and the approximate nature of this relation, we conservatively adopt 
a $1\sigma$ upper limit of $\Delta t \simeq 1$ Gyr, and translate this
timescale into a specific SFR 
$(\text{sSFR} \equiv \text{SFR}/M) = 1/\Delta t \simeq 1\,\mathrm{Gyr}^{-1}$.
At the median stellar mass of our
sample, the corresponding lower limit for the SFR just prior to
the quenching is therefore $\text{SFR}\simeq 200 \,M_\odot\,\text{yr}^{-1}$.
This matches the measured SFR of main-sequence 
star-forming galaxies of similar stellar mass at $z\gtrsim2$ 
\citep[e.g.,][]{daddi:2007:sfr,pannella:2014,rodighiero:2014}.

As an independent  consistency check we have adopted 
an exponentially rising SFR as appropriate for star-forming main-sequence
galaxies at $z \sim 2$
\citep[e.g.,][]{renzini:2009,maraston:2010,papovich:2011,reddy:2012}
and have estimated 
that a SFR culminating at $\sim 300\,M_\odot\,\text{yr}^{-1}$
just prior to the quenching would be required to build a stellar mass of
$2.3\times10^{11}\,M_\odot$ by $z=2.3$,
i.e., at the average quenching epoch and mass for our sample.
Again, this is in good agreement with the
pre-quenching SFR that we have inferred from the \afe ratio.

Detailed studies of massive, $M>10^{11} M_\odot$ star-forming main-sequence galaxies at $z>2$--$2.5$ 
with resolved kpc-scale gas kinematics, SFR and stellar mass densities, and emission line profiles 
secured in recent years 
(\citealt{forsterschreiber:2009}, and N.~M.~F\"orster~Schreiber et al. 2015, in preparation; \citealt{tacchella:2015:data}) 
show that their high SFRs 
are mostly sustained at high galactocentric distances 
\citep{genzel:2014a, tacchella:2015:science}. 
By  contrast, the central regions of such galaxies are already
relatively quenched, and have reached central mass concentrations that
are similar to those of $z=0$ quenched spheroids \citep{tacchella:2015:science}.
These inner quenched spheroidal components argue for an inside-out
quenching process, as expected, for example, in the case of AGN feedback or
gravitational quenching \cite[but see][]{sargent:2015}.
Moreover, these galaxies show strong nuclear
outflows running at up to $\sim 1600\,\kms$, which are most likely driven
by AGN feedback
\citep{cimatti:2013,forsterschreiber:2014,genzel:2014b}.  These outflows
provide tantalizing support for AGN-related processes being responsible for quenching, though evidence for AGN
feedback is not necessarily evidence for AGN quenching.

These $z\gtrsim 2$ galaxies qualify as possible precursors to our quenched galaxies at
$z\sim 1.6$ for two reasons. First, all $z\sim 2.3$ massive galaxies on the
main sequence must soon be quenched, otherwise they would dramatically overgrow \citep{renzini:2009,peng:2010}.
Second, their properties (described above) indicate that the quenching process may already be under way.
These observations seem to support the notion that \textit{mass quenching} of star formation in massive galaxies
\citep{peng:2010} is predominantly a rapid process,
as opposed to an slow quenching process such as halo quenching
\citep[][but see \citealt{knobel:2015}]{woo:2015}.

\subsection{Size growth and progenitor bias}

As we showed above, a purely passive evolution of our quenched galaxies to $z\sim0$, i.e., without any further star formation, 
will bring their stellar population properties to closely match their counterparts in the local Universe. 
However, quenched galaxies at high redshifts are known to have smaller sizes  compared to local galaxies 
\citep[e.g.,][]{daddi:2005:pbzk,trujillo:2006,cimatti:2008,vandokkum:2008,cassata:2011,vanderwel:2014}. 
The average effective radius of 2.7 kpc that we have measured for the 24 $\mz=1.6$ galaxies is, indeed, 
a factor of $\sim 2.5$ smaller than that of local quenched galaxies of the same stellar mass 
(\autoref{tab:objprop}; \citealt{newman:2012}).  
Two main processes  have been suggested and extensively discussed being responsible for the average size growth of the quenched galaxy population: 
the growth of individual galaxies and the addition of larger-sized quenched galaxies at later epochs \citep{carollo:2013}. 
A series of minor {\it dry} mergers with other quenched galaxies is one effective way to increase the sizes of individual galaxies, by adding an extended envelope to a pre-existing compact galaxy
\citep[e.g.,][]{hopkins:2009b,naab:2009,cappellari:2013a}. 
Based on the study of abundance ratios of different elements,
\citet{greene:2012,greene:2013} found that stellar populations in the outskirts ($\gtrsim 2r_\text{e}$) 
of nearby massive early-type galaxies ($\sigma_\star\gtrsim 150\,\kms$) 
are different from those in the inner regions, as they are composed of similarly old ($\sim 10$ Gyr) but more metal-poor ($\feh \sim -0.5$)
and $\alpha$-element enhanced ($\afe \sim 0.3$) stars. 
They argue that these distinct stellar populations have formed at $z>1.5$--$2$ in less massive systems 
and then accreted onto the outskirts of massive quenched galaxies. 
This scenario is consistent with our results, as the stellar population parameters of both low and high-redshift galaxies considered here
were measured essentially within the central part of the galaxies.

One problem with this mechanism for growing galaxies is that it requires minor merging events to be gas-poor,
otherwise gas would sink to the bottom of the potential well, leading to further star formation and resulting in an even more compact galaxy \citep{naab:2009}.
However, at $z\gtrsim1.5$ most galaxies appear to be gas-rich and forming stars \citep[e.g.,][]{daddi:2010:co, tacconi:2010, tacconi:2013, sargent:2014},
and it seems  rather difficult for massive quenched galaxies 
to accrete selectively quenched galaxies 
and  avoid gas accretion and subsequent star formation.  
Still, one may speculate that \textit{galactic conformity},
the phenomenon that satellites of quenched centrals are more likely to be quenched even out to $z\lesssim 2$
\citep[e.g.,][]{weinmann:2006, hartley:2014, knobel:2015}, might help make small systems quiescent before they are accreted.
Moreover, \citet{williams:2011} and \citet{newman:2012} do not find a sufficient number of satellites
around massive quenched galaxies to account for their size growth between $z=2$ and $1$.

Thus an alternative or additional process to secularly grow the average size of quenched galaxies needs to be considered.
At stellar masses $\gtrsim 10^{11}\,M_\odot$, 
the number density of quenched galaxies increases by about one order of magnitude between $z=2$ and $z=0$ (e.g., \citealt{cassata:2013}; \citealt{muzzin:2013:mf}), 
with most of the increase for massive galaxies having taken place by $z \sim 1$ \citep[e.g.,][]{cimatti:2006}. Thus, it has been argued that at least part of the evolution in the average size of quenched galaxies should be due to the continuous addition of quenched galaxies that are already larger than those formed at earlier times
\citep[e.g.,][]{carollo:2013,poggianti:2013}.  Indeed, the average size of star-forming galaxies also increases with time, running above and nearly parallel to the average size of quenched galaxies  \citep{vanderwel:2014}. Therefore, newly quenched galaxies are likely to be larger than pre-existing ones and also smaller than the star-forming progenitors 
due to the fast fading of the quenched disk \citep[][]{carollo:2014}.
This effect, a kind of \textit{progenitor bias}, could explain some of the apparent size evolution of quenched galaxy populations,
given the mentioned difficulties of the scenario invoking the size growth of individual galaxies,
though the two effects are also likely to work together \citep[][]{belli:2015}.

For these reasons, a comparison between galaxy populations at different redshifts and fixed  stellar masses
could suffer from this progenitor bias. However, \citet{bezanson:2012} showed that the number density of quiescent galaxies 
with large $\sigma_\text{inf}$ ($\gtrsim 250\,\kms$) is remarkably constant at $0.3<z<1.5$, 
which indicates that the build-up of quiescent galaxy populations 
with high  velocity dispersions was largely completed by $z\sim1.5$. 
Thus, a comparison based on stellar velocity dispersions 
could greatly reduce the effect of the progenitor bias.
Indeed, in \autoref{fig:pop_sigmaz} essentially all of the nearby early-type galaxies presented in 
\citet{spolaor:2010} with $\sigma_\star>250$ km s$^{-1}$ have stellar population ages  $\gtrsim 10$ Gyr.  
The age distribution of morphologically selected SDSS early-type galaxies 
with $\sigma_\star\gtrsim 250$  km s$^{-1}$ at $0.05<z<0.06$ also peaks at an age of $\simeq$ 10 Gyr \citep{thomas:2010},
which is in good agreement with the expected age for our $\mz=1.6$ sample 
assuming pure passive evolution down to $z \sim 0.05$.

%%%%%%%%%%%%%%%%%%%%%%%%%%%%%%%%%%%%%%%%%%%%%%%%%%%%%%%%%%%
\section{Summary}
\label{sec:summary}

Using a rest-frame optical composite spectrum of 24 massive quenched galaxies at $\mz=1.6$ and with $\mm=2.5\times 10^{11}\, M_\odot$, 
we have derived luminosity-weighted  stellar population parameters from Lick absorption line indices, namely age, \zoh, and \afe, and 
have discussed their past SFH  and subsequent evolution toward lower redshifts, identifying the likely progenitors and descendants of similar galaxies.
Our main results can be summarized as follows. 

\begin{itemize}
\item The average stellar population properties derived are 
  $\log(\text{age}/\text{Gyr})=0.04_{-0.08}^{+0.10}$,  or $\sim 1.1$ Gyr, 
  $\zoh=0.24_{-0.14}^{+0.20}$, and $\afe=0.31_{-0.12}^{+0.12}$.
  The average stellar velocity dispersion of  these galaxies is $\sigma_\star=330 \pm 41$ \kms, 
  as measured from the composite spectrum.
\item The $\mz=1.6$ galaxies show \zoh and \afe 
  in excellent agreement with those of local early-type galaxies at similar velocity dispersions. 
  Pure passive evolution to $z=0$ brings the age of these $\mz=1.6$ quenched galaxies 
  to coincide with  that of their local counterparts  at the same $\sigma_\star$. 
  Therefore, the stellar populations of the galaxies in our sample qualify such galaxies as plausible progenitors 
  of similarly massive quenched galaxies in the local Universe.
\item The age of 1.1 Gyr for the bulk of stars in these galaxies points to a formation redshift of $z_\text{f} \sim 2.3$,
  an epoch when massive galaxies on the main sequence 
  are very rapidly growing in mass and therefore must soon be quenched.
\item  The \afe value indicates a star formation timescale of $\Delta t \lesssim 1$ Gyr 
  which in turn implies a typical SFR  of $\sim 200\, M_\odot\; {\rm yr}^{-1}$  just prior to the quenching.
  This SFR is well within a range covered by similarly massive main-sequence galaxies at $z \sim 2.3$.
\item These properties of massive $z \sim 2.3$ main-sequence galaxies 
  qualify them as likely precursors to our quenched galaxy at $\mz=1.6$.
\end{itemize}

One limitation of the present study is that we had to stack individual spectra to an equivalent integration time of $\sim 200$ hr in order
to achieve a high S/N that was adequate for the analysis of the Lick indices. 
Further studies based on  high S/N spectra of individual galaxies are needed to account for 
a diversity in high-redshift quenched galaxy populations \citep[e.g.,][]{belli:2015}.
This is becoming possible through deep absorption line spectroscopy 
with the new generation of near-IR spectrographs such as Keck/MOSFIRE and VLT/KMOS, though very long integration times will still be necessary.

%%%%%%%%%%%%%%%%%%%%%%%%%%%%%%%%%%%%%%%%%%%%%%%%%%%%%%%%%%%
%%%%%%%%%%%%%%%%%%%%%%%%%%%%%%%%%%%%%%%%%%%%%%%%%%%%%%%%%%%
%%%%%%%%%%%%%%%%%%%%%%%%%%%%%%%%%%%%%%%%%%%%%%%%%%%%%%%%%%%
%%%%%%%%%%%%%%%%%%%%%%%%%%%%%%%%%%%%%%%%%%%%%%%%%%%%%%%%%%%

\acknowledgments
We are grateful to the referee for useful and constructive comments. 
We thank the staff of the Subaru telescope, especially Ichi Tanaka, 
for supporting the observations. We also thank Sumire Tatehora, 
Will Hartley, Claudia Maraston, Daniel Thomas, 
and Alexandre Vazdekis for fruitful discussions,
and Joanna Woo for reading the manuscript thoughtfully. 
We acknowledge Inger J{\o}rgensen for providing us with their measurements 
in electronic form. 
This work has been partially supported by the Grant-in-Aid 
for the Scientific Research Fund under Grant No.23224005, 
and the Program for Leading Graduate Schools PhD Professional: 
Gateway to Success in Frontier Asia, commissioned 
by the Ministry of Education, Culture, 
Sports, Science and Technology (MEXT) of Japan. 
This research made extensive use of \texttt{Astropy}\footnote{\url{http://www.astropy.org}}, 
a community-developed core Python package for Astronomy \citep{astropy},
and \texttt{matplotlib}\footnote{\url{http://matplotlib.org}}, a Python 2D plotting library \citep{matplotlib}.

{\it Facilities:} \facility{Subaru (MOIRCS)}

%% \bibliography{bibliography}

\end{document}